\documentclass[a4paper,12pt]{article}
\usepackage{amssymb}
\usepackage{mathrsfs}
\usepackage{cite}

\csname @addtoreset\endcsname{equation}{section}

\newcommand{\dd}{\mathrm{d}}
\newcommand{\tr}{\mathrm{Tr}\;}
\newcommand{\sder}{\mathcal{D}} 
\newcommand{\sderl}{\overleftarrow{\mathcal{D}}}

\newcommand{\p}{\partial}
\newcommand{\partiall}{\overleftarrow\partial}

\newcommand{\nablal}{\overleftarrow\nabla}
\newcommand{\nablar}{\overrightarrow\nabla}
\newcommand{\lie}{\mathcal{L}}
\newcommand{\dual}{{}^*}

\newcommand{\der}[2]{{\partial #1\over\partial #2}}

\newcommand{\eq}{\begin{equation}}
\newcommand{\feq}{\end{equation}}
\newcommand{\eqn}{\begin{eqnarray}}
\newcommand{\feqn}{\end{eqnarray}}
\newcommand{\arr}{\begin{eqnarray*}}
\newcommand{\farr}{\end{eqnarray*}}

\def\al{\alpha}
\def\be{\beta}
\def\ga{\gamma}
\def\de{\delta}
\def\ep{\epsilon}
\def\eps{\varepsilon}
\def\ze{\zeta}

\def\si{\sigma}
\def\om{\omega}

\def\Ga{\Gamma}

\def\Om{\Omega}

\begin{document}

\begin{titlepage}
\begin{flushright}
IFUM-763-FT\\
hep-th/0307022
\end{flushright}
\vspace{.3cm}
\begin{center}
\renewcommand{\thefootnote}{\fnsymbol{footnote}}
{\Large \bf All Supersymmetric Solutions of $N = 2$, $D = 4$ Gauged Supergravity}
\vskip 25mm
{\large \bf {Marco M.~Caldarelli\footnote{marco.caldarelli@mi.infn.it}
and Dietmar Klemm\footnote{dietmar.klemm@mi.infn.it}}}\\
\renewcommand{\thefootnote}{\arabic{footnote}}
\setcounter{footnote}{0}
\vskip 10mm
{\small
Dipartimento di Fisica dell'Universit\`a di Milano\\
and\\
INFN, Sezione di Milano,\\
Via Celoria 16, I-20133 Milano.\\
}
\end{center}
\vspace{2cm}
\begin{center}
{\bf Abstract}
\end{center}
{\small We classify all supersymmetric solutions of minimal gauged supergravity
in four dimensions. There are two classes of solutions that are distinguished
by the norm of the Killing vector constructed from the Killing spinor.
If the Killing vector is timelike, the solutions are determined by the
geometry of a two-dimensional base-manifold. When it is lightlike, the
most general BPS solution is given by an electrovac AdS travelling wave.
This supersymmetric configuration was previously unknown.
Generically the solutions preserve one quarter of the supersymmetry.
Also in the timelike case we show that there exist new BPS solutions, which
are of Petrov type I, and are thus more general than the previously known
type D configurations. These geometries can be uplifted to obtain new solutions
of eleven-dimensional supergravity.
}

\end{titlepage}


\section{Introduction}

\label{intro}

Supersymmetric solutions to supergravity theories have played, and continue to
play, an important role in developments in string theory. This makes it
desirable to obtain a complete classification of BPS solutions to various
supergravities in diverse dimensions. Progress in this direction has been made
recently using the mathematical notion of G-structures
\cite{Gauntlett:2002sc}.
The basic strategy is
to assume the existence of at least one Killing spinor, and to construct
differential forms as bilinears from this Killing spinor. These forms, which
define a preferred G-structure, obey several algebraic and differential
equations that can be used to deduce the metric and the other bosonic
supergravity fields.  Despite some partial progress
\cite{Gauntlett:2002fz,Gauntlett:2003cy}, a complete classification of
supersymmetric geometries in eleven- and ten-dimensional supergravities seems
to be currently out of reach.
This motivates to consider
simpler, lower-dimensional supergravities, in particular the ones that can be
obtained by Kaluza-Klein reduction from ten or eleven dimensions. A systematic
classification of supersymmetric solutions has been obtained for minimal
supergravity in five dimensions, both in the ungauged \cite{Gauntlett:2002nw}
and gauged case \cite{Gauntlett:2003fk}, and for minimal supergravity in six
dimensions \cite{Gutowski:2003rg}.

Here we present a complete classification of BPS solutions in minimal gauged
four-dimensional supergravity. Using the techniques of \cite{Chamblin:1999tk}
to uplift the geometries to eleven dimensions, our analysis can be viewed as a
classification of a restricted class of eleven-dimensional solutions.
The most general configuration admitting Killing spinors in the ungauged
four-dimensional case was obtained by Tod many years ago \cite{Tod:pm},
generalizing earlier work by Gibbons and Hull \cite{Gibbons:fy}. In the absence
of dust sources, the resulting geometry is given by an Israel-Wilson-Perjes
metric if the Killing vector constructed from the Killing spinor is timelike,
and by a plane-wave spacetime if this Killing vector is lightlike. In the
gauged case, we will find more complicated BPS solutions: In the timelike case,
they are determined by the geometry of a two-dimensional base manifold, whereas
in the lightlike case they are given by electrovac AdS travelling waves.  A
further new feature is the fact that generically the solutions preserve only
one quarter of the supersymmetry, whereas in the ungauged case they are one
half supersymmetric \cite{Tod:pm}.

Supersymmetric solutions to minimal gauged supergravity in four dimensions have
been studied in
\cite{Romans:1991nq,Kostelecky:1995ei,Caldarelli:1998hg,Cacciatori:1999rp,Alonso-Alberca:2000cs,Brecher:2000pa}.
Among these are the one quarter supersymmetric magnetic monopoles
\cite{Romans:1991nq,Caldarelli:1998hg} that have no well-defined limit when the
gauge coupling constant goes to zero, because their magnetic charge is
quantized in terms of the inverse coupling constant. Apart from the pp-waves in
AdS considered in \cite{Brecher:2000pa}, which have Petrov type N, the BPS
solutions that were hitherto known do all belong to a subclass of the most
general Petrov type D metric found by Pleba\'nski and Demia\'nski
\cite{Plebanski:gy}, which is determined by electric and magnetic charges, nut
charge, and mass and rotation parameters.  The conditions under which this
metric admits Killing spinors were obtained in \cite{Alonso-Alberca:2000cs}. We
shall use our formalism to show that there exist also BPS solutions that have
Petrov type I, and therefore do not belong to the classes mentioned above.

The remainder of this paper is organized as follows: In section \ref{sugra}, we
briefly review minimal gauged supergravity in four dimensions, and obtain the
algebraic and differential constraints satisfied by the differential forms
constructed as bilinears in the Killing spinor. In section \ref{timelike} these
constraints are solved for the case in which the Killing vector resulting from
the Killing spinor is timelike, and the metric and electromagnetic field
strength are deduced. As specific examples of the general BPS solution, we
recover the one quarter supersymmetric Reissner-Nordstr\"om-Taub-Nut-AdS
spacetime obtained in \cite{Alonso-Alberca:2000cs}, and construct a previously
unknown supersymmetric Petrov type I configuration.  In section
\ref{lightlike}, the lightlike case is analyzed, and it is shown that the most
general supersymmetric geometry is given by an electrovac AdS travelling
wave. In section \ref{maxsusy} we show that the only maximally supersymmetric
solution is AdS$_4$ with vanishing gauge field.
We conclude with some final remarks in section \ref{final}. The appendix
contains our conventions and some useful identities.


\section{$N=2$, $D=4$ Gauged Supergravity} \label{sugra}
The gauged version of $N=2$ supergravity was found by Freedman and
Das \cite{Freedman:1976aw} and by Fradkin and Vasiliev
\cite{Fradkin:1976xz}. In this theory, the rigid $\mathrm{SO(2)}$ symmetry
rotating the two independent Majorana supersymmetries present in the ungauged
theory, is made local by introduction of a minimal gauge coupling between the
photons and the gravitini. Local supersymmetry then requires a negative
cosmological constant and a gravitini mass term.
The theory has four bosonic and four fermionic degrees of freedom;
it describes a graviton $e_\mu^a$, two real gravitini $\psi_\mu^i$ $(i=1,2)$,
and a Maxwell gauge field ${\mathcal A}_{\mu}$. As we said, the latter is
minimally coupled to the gravitini, with coupling constant $\ell^{-1}$. If we
combine the two $\psi_\mu^i$ to a single complex spinor
$\psi_\mu=\psi_\mu^1+i\psi_\mu^2$, the Lagrangian reads
(cf.~also \cite{Romans:1991nq})
\eqn
e^{-1}L &=& -\frac{1}{4}R
            + \frac{1}{4}{\mathcal F}_{\mu\nu}{\mathcal F}^{\mu\nu}
            - \frac{3}{2\ell^2}
            + \frac{1}{2}\bar{\psi}_\mu\Ga^{\mu\nu\rho} D_\nu\psi_\rho
	     \nonumber \\
	&& \qquad
            + \frac{i}{8}({\mathcal F}^{\mu\nu} + \hat{{\mathcal F}}^{\mu\nu})
            \bar{\psi}_\rho\Ga_{\left[\mu\right.}\Ga^{\rho\si}
	           \Ga_{\left.\nu\right]}
            \psi_\si
            - \frac{1}{2\ell}\bar{\psi}_\mu\Ga^{\mu\nu}\psi_\nu\,.
            \label{lagrange}
\feqn
We see that the cosmological constant is $\Lambda = -3\ell^{-2}$.
$D_\mu$ denotes the gauge- and Lorentz-covariant derivative defined by
\begin{equation}
D_\mu = \nabla_\mu - i\ell^{-1}{\mathcal A}_\mu\,, \label{gaugecovder}
\end{equation}
where $\nabla_\mu$ is the Lorentz-covariant derivative
\begin{equation}
\nabla_\mu = \partial_\mu + \frac{1}{4}\om_\mu^{\;\;ab}\Ga_{ab}\,.
\end{equation}
The equation of motion for the spin connection $\om_\mu^{\;\;ab}$ reads
\begin{equation}
\om_{\mu ab} = \Om_{\mu ab} - \Om_{\mu ba} - \Om_{ab\mu}\,,
\end{equation}
where
\begin{equation}
\Om_{\mu\nu}^{\;\;\;\;a} \equiv \partial_{\left[\mu\right.}
             e_{\left.\nu\right]}^{\;a}
                    - \frac{1}{2}{\mathrm{Re}}(\bar{\psi}_\mu\Ga^a\psi_\nu)\,.
\end{equation}
$\hat{\mathcal F}_{\mu\nu}$ denotes the supercovariant field strength given by
\begin{equation}
  \hat{\mathcal F}_{\mu\nu} = {\mathcal F}_{\mu\nu}
  - {\mathrm{Im}}(\bar{\psi}_\mu\psi_\nu)\,.
\label{supcovfs}
\end{equation}
The action is invariant under the local supersymmetry transformations
\eqn
\de e_\mu^{a} &=& {\mathrm{Re}}(\bar{\epsilon}\Ga^a\psi_\mu)\,, \nonumber \\
\de{\mathcal A}_\mu &=& {\mathrm{Im}}(\bar{\epsilon}\psi_\mu)\,,
        \label{transfsusy}\\
\de \psi_\mu &=& \sder_\mu \epsilon\,. \nonumber 
\feqn
In (\ref{transfsusy}) $\epsilon$ is an infinitesimal Dirac spinor, and
$\sder_\mu$ is the supercovariant derivative defined by
\begin{equation}
\sder_\mu = D_\mu + \frac{1}{2\ell}\Ga_\mu 
+ \frac{i}{4}\hat{\mathcal F}_{ab}\Ga^{ab}
                 \Ga_\mu\,. \label{supcovder}
\end{equation}
Invariance of bosonic backgrounds under the local supersymmetry transformations
(\ref{transfsusy}) yields the Killing spinor equation
\begin{equation}
\sder_\mu \epsilon = 0\,. \label{killspinequ}
\end{equation}

\subsection{Differential Forms Constructed from the Killing Spinor}

Any supersymmetric geometry must admit at least one Killing spinor $\epsilon$,
from which we can construct the following bosonic differential forms:
\begin{itemize}
  \item[-] a scalar $f=\bar\ep\ep$\,,
  \item[-] a pseudo-scalar $g=i\bar\ep\Ga_5\ep$\,,
  \item[-] a vector $V_\mu=i\bar\ep\Ga_\mu\ep$\,,
  \item[-] a pseudo-vector $A_\mu=i\bar\ep\Ga_5\Ga_\mu\ep$\,,
  \item[-] an antisymmetric tensor $\Phi_{\mu\nu}=i\bar\ep\Ga_{\mu\nu}\ep$\,.
\end{itemize}
The $i$ factors have been inserted in such a way that all these
quantities are real.
As the matrices $\mathbf{1}$, $\Ga_a$, $\Ga_5\Ga_a$ and $\Ga_{ab}$
form a basis of the space of $4\times4$ matrices, any other differential form
constructed from the spinor $\ep$ can be written as a linear
combination of the previous ones.\\
In all, we have 16 real components, which cannot be all
independent since an arbitrary spinor has real dimension 8. There must thus
exist some algebraic relations between the differential forms, which can be
obtained using the Fierz identity.
From the spinor $\ep$, we can construct a matrix $\ep\bar\ep$, which
has as components in the $\Ga$-basis\footnote{
  Any matrix $M$ can be expanded as
  $4M=\mathbf{1}\cdot\tr M
  +\Ga_\mu\tr\left(M\Ga^\mu\right)
  -\frac12\Ga_{\mu\nu}\tr\left(M\Ga^{\mu\nu}\right)
  -\Ga_5\Ga_\mu\tr\left(M\Ga_5\Ga^\mu\right)
  +\Ga_5\tr\left(M\Ga_5\right)$. Taking $M=\ep\bar\ep$ and using
  $\tr\ep K\bar\ep=\bar\ep K\ep$ for any matrix $K$ we obtain the Fierz
  identity.
}
\begin{equation}
  4\ep\bar\ep=f\cdot\mathbf{1}-iV^\mu\Ga_\mu+\frac i2\Phi^{\mu\nu}\Ga_{\mu\nu}
  +iA^\mu\Ga_5\Ga_\mu-ig\Ga_5\,.
\end{equation}
As a consequence,
\begin{equation}
  iV^\mu\Ga_\mu\ep=-iA^\mu\Ga_5\Ga_\mu\ep=-\left(f+ig\Ga_5\right)\ep
\label{fierz}\end{equation}
and
\begin{equation}
  i\Phi^{\al\be}\Ga_{\al\be}\ep=2\left(f-ig\Ga_5\right)\ep\,.
\end{equation}
Contracting these relations with $\bar\ep$ we obtain
\begin{equation}
  N = V^2=-A^2=-\left(f^2+g^2\right),\qquad
  f^2-g^2=\frac12\Phi^2\,,
\label{N}\end{equation}
where we have defined $N$ as the norm squared of $V$. Hence $N\leq0$ and it
follows that $V$ is either timelike (and $A$ spacelike) or lightlike.
A contraction with $\bar\ep\Ga_5$ yields $V\cdot A=0$.
The other relations which can be found from the Fierz identity are
\begin{equation}
  fV_\mu=\frac12\eps_{\mu\nu\rho\si}A^\nu\Phi^{\rho\si}\,, \qquad
  fA_\mu=\frac12\eps_{\mu\nu\rho\si}V^\nu\Phi^{\rho\si}\,, \label{fVfA}
\end{equation}
\begin{equation}
  gV_\mu=\Phi_{\mu\nu}A^\nu\,, \qquad
  gA_\mu=\Phi_{\mu\nu}V^\nu\,, \label{gVgA}
\end{equation}
\begin{equation}
  fg=-\frac18\eps_{\mu\nu\rho\si}\Phi^{\mu\nu}\Phi^{\rho\si}\,,
     \label{fg}
\end{equation}
\begin{equation}
  f\Phi_{\mu\nu}=-\eps_{\mu\nu\rho\si}V^\rho A^\si
  +\frac12g\eps_{\mu\nu\rho\si}\Phi^{\rho\si}\,,
  \label{fPhi}
\end{equation}
\begin{equation}
\Phi_{(\mu}{}^\rho\eps_{\nu)\rho\al\be}\Phi^{\al\be}
-\frac14g_{\mu\nu}\eps_{\rho\si\al\be}\Phi^{\rho\si}\Phi^{\al\be}=0\,.
\label{PhiPhi}
\end{equation}

After having obtained the algebraic relations obeyed by the differential
forms, let us now derive the differential constraints.
Starting from the supercovariant derivative (\ref{supcovder}), one can define
a left-acting supercovariant derivative
$\sderl_\mu$ by the relation
$\bar\psi\ \sderl_\mu\equiv\overline{\sder_\mu\psi}$, yielding
\begin{equation}
  \sderl_\mu = \nablal_\mu+\frac i\ell{\mathcal A}_\mu - \frac1{2\ell}\Ga_\mu -
  \frac{i}{4}{\mathcal F}_{ab}\Ga_\mu\Ga^{ab}\,,
\end{equation}
where we have defined the left-acting covariant derivative
\begin{equation}
\nablal\equiv\partiall-\frac14\omega_\mu{}^{ab}\Ga_{ab}\,,
\end{equation}
such that $\bar\psi\ \nablal_\mu\equiv\overline{\nabla_\mu\psi}$.
With these definitions, we obtain a Leibniz rule for the
derivative of the contraction of two spinors
\begin{equation}
  \partial_\mu\left(\bar\psi\chi\right)=\bar\psi\nablal_\mu\chi+
\bar\psi\nablar_\mu\chi\,.
\end{equation}
With some algebra the previous identity can be generalized for
insertions of $\Gamma$-matrices:
\begin{equation}
  \nabla_\mu\left(\bar\psi\Ga_{\mu_1}\cdots\Ga_{\mu_n}\chi\right)
  =\bar\psi\nablal_\mu\Ga_{\mu_1}\cdots\Ga_{\mu_n}\chi+
  \bar\psi\Ga_{\mu_1}\cdots\Ga_{\mu_n}\nablar_\mu\chi\,,
\end{equation}
and with a $\Ga_5$ matrix:
\begin{equation}
  \nabla_\mu\left(\bar\psi\Ga_5\Ga_{\mu_1}\cdots\Ga_{\mu_n}\chi\right)
  =\bar\psi\nablal_\mu\Ga_5\Ga_{\mu_1}\cdots\Ga_{\mu_n}\chi+
  \bar\psi\Ga_5\Ga_{\mu_1}\cdots\Ga_{\mu_n}\nablar_\mu\chi\,.
\end{equation}
Using these relations as well as $\sder_\mu \epsilon = 0$, it is
straightforward to show that $f$, $g$, $V_\mu$, $A_\mu$ and $\Phi_{\mu\nu}$
satisfy the differential constraints
\begin{equation}
\partial_\mu f={\mathcal F}_{\mu\nu}V^\nu\,,
\label{eqf}\end{equation}
\begin{equation}
  \partial_\mu g=-\frac1\ell A_\mu-\frac12\eps_{\mu\nu\rho\si}V^\nu
  {\mathcal F}^{\rho\si}\,,
\label{eqg}\end{equation}
\begin{equation}
  \nabla_\mu V_\nu=\frac1\ell\Phi_{\mu\nu}-f{\mathcal F}_{\mu\nu}
  +\frac12g\eps_{\mu\nu\rho\si} {\mathcal F}^{\rho\si}\,,
\label{eqV}\end{equation}
\begin{equation}
  \nabla_\mu A_\nu=-\frac1\ell gg_{\mu\nu}
  -{\mathcal F}_{(\mu}{}^\rho\eps_{\nu)\rho\al\be}\Phi^{\al\be} 
  +\frac14g_{\mu\nu}\eps_{\rho\si\al\be}{\mathcal F}^{\rho\si}\Phi^{\al\be}\,,
\label{eqA}\end{equation}
\begin{equation}
  \nabla_\mu\Phi_{\al\be}=\frac2\ell g_{\mu[\al}V_{\be]}
  +2{\mathcal F}_{[\al}{}^\si\eps_{\be]\si\mu\nu}A^\nu
  +{\mathcal F}_\mu{}^\si\eps_{\si\al\be\nu}A^\nu
  +g_{\mu[\al}\eps_{\be]\nu\rho\si}A^\nu{\mathcal F}^{\rho\si}\,.
\label{eqPhi}\end{equation}

Furthermore, taking the symmetric part of (\ref{eqV}) we
obtain $\nabla_{(\mu}V_{\nu)}=0$, hence $V^\mu$ is a Killing vector.


\section{The Timelike Case}\label{timelike}

\subsection{Construction of the Bosonic Fields}
\label{constrbos}

In the timelike case we have $N<0$, hence $V$ is a timelike vector and $A$ a
spacelike vector. Furthermore, the antisymmetric part of (\ref{eqA}) implies
$\dd A=0$, so that locally there exists a function $z$ such that $A=\dd z$.
Introducing the coordinates $(t,x,y,z)$ such that $V = \partial_t$, and $x, y$
are two transverse coordinates, the metric can be written as
\begin{equation}
  \dd s^2=N\left(\dd t+\om\right)^2-\frac1N\left(\,\dd z^2
  +h_{ij}\,(\dd x^i + a^i \dd z)(\dd x^j + a^j \dd z)\right)\,.
\end{equation}
Here the latin indices $i,j,\ldots$ range from 1 to 2, with $x^1=x$,
$x^2=y$. As $V$ is Killing, the function $N$, the one-form $\omega$,
the transverse metric $h_{ij}$ as well as $a^i$ are functions of $(x,y,z)$ but
are independent of $t$; the metric is stationary.
A contraction of $V$ with equations (\ref{eqf}) and (\ref{eqg}) shows that the
functions $f$ and $g$ are also time-independent.
Note that $a^i$ can be eliminated by a diffeomorphism
\begin{equation}
x^i = x^i({x'}^j, z)\,, \qquad {\mathrm{with}} \qquad
      \frac{\partial x^i}{\partial z} = - a^i\,,
\end{equation}
so that we can set $a^i = 0$ without loss of generality.
The resulting metric is invariant under the transformation
\begin{equation}
t\mapsto t+\chi(x,y,z)\,,\qquad
\om\mapsto\om-\dd\chi
\label{gaugeom}\end{equation}
where $\chi$ is an arbitrary function of $x$, $y$ and $z$. We can use this
freedom to eliminate the $\om_z$ component of the shift vector. Hence, without
loss of generality, we can take $\om=\om_i\,\dd x^i$.

We come now to the two-form $\Phi$ and the electromagnetic field strength
$\mathcal F$.
Following appendix~\ref{decomp}, we decompose $\Phi$ with respect to the vector
$V$ and, using equations (\ref{fVfA}) and (\ref{gVgA}), we obtain
\begin{equation}
\Phi=-\frac1N\left[gV\wedge A-f\dual\left(V\wedge A\right)\right]\,.
\label{Phi}\end{equation}
Similarly, equations (\ref{eqf}), (\ref{eqg}) immediately yield the
decomposition of $\mathcal F$ with respect to the Killing vector $V$,
\begin{equation}
{\mathcal F}=-\frac1N\left[V\wedge\dd f+\dual\left(V\wedge\left(\dd
  g+\frac1\ell A\right)\right)\right]\,.
\label{F}\end{equation}
The electromagnetic field is invariant under the 1-parameter group of
transformations generated by the timelike Killing vector $V$
\begin{equation}
\lie_V{\mathcal F}=\lie_V\dual{\mathcal F}=0\,,
\end{equation}
hence $\mathcal F$ is a stationary electromagnetic field, and the full solution
is invariant under the time translations generated by $V$.

In order to determine the shift functions $\omega_i$, we first note that
$V=N\left(\dd t+\omega\right)$ as a
one-form. Combining its exterior derivative
$\dd V=\frac{\dd N}N\wedge V+\dd\omega$ with the equation (\ref{eqV}) for $V$
and substituting $\Phi$ and $\mathcal F$ using equations (\ref{Phi}) and
(\ref{F}), we obtain
\begin{equation}
\dd\omega=\frac2{N^2}\phantom{\frac12}
\dual\left[V\wedge\left(f\dd g-g\dd f+\frac2\ell fA\right)\right]\,,
\end{equation}
or, in components,
\begin{equation}
\p_z\om_i=\frac{2\sqrt h}{N^2}h^{jl}\epsilon_{il}\left(f\p_jg-g\p_jf\right)\,,
\label{omegaz}\end{equation}
\begin{equation}
\p_i\om_j-\p_j\om_i=
\frac{2\sqrt{h}}{N^2}\epsilon_{ij}\left(f\p_zg-g\p_zf+\frac{2f}\ell\right)\,,
\label{omegaij}\end{equation}
where $h=\det h_{ij}$ and $\epsilon_{12} = 1$.
If the functions $f$ and $g$ are known, these equations
completely determine $\om_i$, up to gauge transformations of the form
(\ref{gaugeom}).

Next we consider equation (\ref{eqA}). We first use the expressions
(\ref{Phi}) and (\ref{F}) for $\Phi$ and $\mathcal F$ in (\ref{eqA})
and project the resulting equation on $A^{\mu}$. In this way one gets
\begin{equation}
\nabla_A A_\nu = -\frac 12 \nabla_\nu N\,,
\end{equation}
which is already satisfied since $A^2 = -N$ and $\dd A = 0$. Similarly,
also the projection on $V^{\mu}$ does not yield anything new, if the
Eqns.~(\ref{gVgA}), (\ref{eqV}), (\ref{F}) as well as $A\cdot V = 0$ hold.
The only nontrivial information comes from the projection on the
transverse metric
\begin{equation}
h_{\mu\nu} = -N g_{\mu\nu} + V_\mu V_\nu - A_\mu A_\nu\,,
\end{equation}
from which we obtain
\begin{equation}
  {h^{\mu}}_{\alpha} {h^{\nu}}_{\beta} \nabla_\mu A_\nu =
  \frac{2gN}{\ell} h_{\alpha\beta}
  + \frac 12 h_{\alpha\beta} A^{\mu} \nabla_\mu N\,. 
  \label{extrcurv}
\end{equation}
This can be viewed as a condition on the extrinsic curvature of the two-surface
with metric $h_{\alpha\beta}$ as embedded into the three-manifold
$\dd z^2 + h_{ij}dx^i dx^j$. Evaluating (\ref{extrcurv}) yields
\begin{equation}
\partial_z h_{ij} = \frac{4g}{\ell N} h_{ij}\,,
\end{equation}
with the general solution
\begin{equation}
h_{ij}=C_{ij}(x^k)\,\exp\left(\int\!\!\frac{4g}{\ell N}\,\dd z\right)\,,
\end{equation}
where $C_{ij}(x^k)$ denotes an arbitrary two-dimensional metric that depends
only on the coordinates $x^k$.

At this point a somewhat lengthy but straightforward calculation shows
that equation (\ref{eqPhi}) for $\Phi$ is
automatically solved.

We finally have to impose the Maxwell equation and Bianchi identity
on $\mathcal F$. It is useful to define the complex variables
\begin{equation}
w=f-ig\,,\qquad
F=\frac{2i}{\ell w}\,,\qquad
\lambda=\exp\int\!\!F\,\dd z\,,
\label{defwF}\end{equation}
in terms of which the combined Maxwell equation and Bianchi identity,
$\dd\left({\mathcal F}+i\dual{\mathcal F}\right)=0$, reduce to
\begin{equation}
\bar\lambda\p^3_z\lambda+\Delta_CF=0\,,
\label{lambda}\end{equation}
where
\begin{equation}
\Delta_C\equiv\frac1{\sqrt{C}}\p_i\left(\sqrt{C}C^{ij}\p_j\right)
=\lambda\bar\lambda\Delta_h
\end{equation}
is the Laplacian of the two-metric $C_{ij}$ and $\Delta_h$ is the Laplacian of
$h_{ij}$. Equivalently, the equation reads
\begin{equation}
\p^3_z\lambda+\lambda\,\Delta_hF=0\,.
\end{equation}
Hence, given an arbitrary two-metric $C_{ij}$, this complex equation determines
the functions $f$ and $g$.

\subsection{Integrability Conditions and Unbroken Supersymmetry Generators}

To see if the solution is indeed invariant under some supersymmetry
transformation, we have to construct the Killing spinor generating it.
A necessary and sufficient condition for its existence (at least locally) is
that the supercurvature vanishes.

First of all, as a consequence of the identities (\ref{fierz}), the Killing
spinor $\ep$ is subject to the constraint $\Ga_{12}\ep=i\ep$, and this
relation already breaks one half of the supersymmetries.

The $t$ component of the Killing spinor equation yields $\p_t\ep=0$,
hence $\ep$ is time-independent. Using (\ref{fierz}), the other supercovariant
derivatives simplify to
\begin{equation}
  \hat\nabla_z\ep=
  \left(\p_z+\frac1{2N}W_+\p_zW_--\frac i\ell{\mathcal A}_z\right)\ep\,,
\end{equation}
\begin{equation}
  \hat\nabla_i\ep=\left(\p_i+\frac1{2N}W_+\p_iW_-
  +\frac i2\check\om^{12}_i-\frac i\ell{\mathcal B}_i\right)\ep\,,
\end{equation}
where we have defined $W_\pm=f\pm ig\Ga_5$, and
\begin{equation}
  {\mathcal B}_i={\mathcal A}_i-f\om_i\,.
\end{equation}
The $W_\pm$ terms can be eliminated by decomposing the spinor $\ep$ into two
chiral spinors $\eta_\pm$ defined by
\begin{equation}
  \eta_+=\frac{\ep_+}{\sqrt{w}}\,,\qquad
  \eta_-=\frac{\ep_-}{\sqrt{\bar w}}\,,\qquad
  \mathrm{with}\qquad
  \ep_\pm=\frac12\left(1\pm\Ga_5\right)\ep\,.
\end{equation}
This leads to
\begin{equation}
  \left(\p_z-\frac i\ell{\mathcal A}_z\right)\eta_\pm=0\quad\mathrm{and}\quad
  \left(\p_i+\frac i2\check\om^{12}_i-\frac i\ell{\mathcal B}_i\right)
  \eta_\pm=0\,. \label{finalkill}
\end{equation}
The identities (\ref{fierz}) tell us that the left- and right-handed spinors
$\eta_-$ and $\eta_+$ are not independent, but related by
$\eta_-=-i\Ga_0\eta_+$.
Together with $\eta_\pm=-i\Ga_{12}\eta_\pm$, we deduce that generically we are
left with only one quarter of the original supersymmetry.

The integrability conditions for (\ref{finalkill}) imply some additional
constraints on the bosonic fields. From the vanishing of the
$(x,y)$ component of the supercurvature one gets
\begin{equation}
  \check{\mathcal R}^{12}=\frac2\ell\check\dd{\mathcal B}\,,
\label{integrability}\end{equation}
where $\check{\mathcal R}^{12}=\check\dd\check\om$ is the curvature form of
the transverse two-manifold with metric $h_{ij}$. For the spin connection
we have thus
\begin{equation}
  \check\om=\frac2\ell{\mathcal B}+\check d\chi\,, \label{twist}
\end{equation}
with $\chi$ an arbitrary function of $x$, $y$ and $z$. (\ref{twist}) is
a ``twisting'' condition, since we can view the coupling to the gauge
field $\mathcal B$ as effectively changing the spin of the supersymmetry
parameter, which becomes a scalar \cite{Maldacena:2000mw}.

The $(z,i)$ component of the integrability conditions yields
\begin{equation}
  \partial_i({\mathcal A}_z+\frac\ell2\p_z\chi)=0\,,
\end{equation}
and determines the function $\chi$ up to an arbitrary function $\psi(z)$,
\begin{equation}
  \chi=-\frac2\ell\int\!\!{\mathcal A}_z\,\dd z+\psi(z)\,.
\end{equation}
Now, the Killing spinor equations reduce to
\begin{equation}
  \left(\p_\mu+\frac i2\p_\mu\left(\chi-\psi\right)\right)\eta_\pm=0\,,
\end{equation}
and are solved by
\begin{equation}
  \eta_\pm=e^{-\frac i2\left(\chi-\psi\right)}\eta^0_\pm
  =e^{\frac i\ell\int\!\!{\mathcal A}_z\,\dd z}\eta^0_\pm\,,
\end{equation}
where $\eta^0_\pm$ obey the constraints $\eta_-=-i\Ga_0\eta_+$ and
$\eta_\pm=-i\Ga_{12}\eta_\pm$. These are satisfied by projecting an
arbitrary spinor $\ep_0$,
\begin{equation}
  \eta_+=\frac14\left(1-i\Ga_{12}\right)\left(1+\Ga_5\right)\ep_0\,,\qquad
  \eta_-=-\frac i4\Ga_0\left(1-i\Ga_{12}\right)\left(1+\Ga_5\right)\ep_0\,.
\end{equation}
Finally, the Killing spinor reads
\begin{equation}
  \ep=\frac14e^{\frac i\ell\int\!\!{\mathcal A}_z\,\dd z}
  \left(\sqrt{f-ig}-i\sqrt{f+ig}\,\Ga_0\right)
  \left(1-i\Ga_{12}\right)\left(1+\Ga_5\right)\ep_0\,.
\end{equation}

We can now state our final result.
Given an arbitrary two-metric $C_{ij}(x^k)$, one obtains the functions $f$
and $g$ by solving equation (\ref{lambda}).
Then, the shift vector $\om$ is obtained as a solution of the equations
(\ref{omegaz}) and (\ref{omegaij}). If in addition the integrability condition
(\ref{twist}) is satisfied, the corresponding field configuration meets all
necessary and sufficient conditions for the existence of the Killing
spinor. In appendix \ref{integrab} it is shown that these conditions, together with
the Maxwell equation and Bianchi identity, imply in the timelike case that also
the Einstein equations hold. In other words, these solutions are BPS states,
preserving generically $1/4$ of supersymmetry.

\subsection{Reissner-Nordstr\"om-Taub-Nut-AdS Solutions}

As an example, we now show how to recover the supersymmetric
Reissner-Nordstr\"om-Taub-Nut-AdS solutions obtained in
\cite{Alonso-Alberca:2000cs}.
Let $C_{ij}$ be a metric of constant curvature $k/l^2$, where without loss of
generality $k = 0, \pm 1$. An explicit form would be for
instance\footnote{$\theta$ and $\phi$ are the coordinates previously
termed $x$ and $y$.}
\begin{equation}
C_{ij} dx^i dx^j = l^2 (d\theta^2 + S_k^2(\theta) d\phi^2)\,, \label{Cij}
\end{equation}
where
\begin{equation}
S_k(\theta) = \left\{\begin{array}{l@{\,,\quad}l}
                     1 & k=0 \\ \sin\theta & k = 1 \\ \sinh\theta & k = -1\,.
                     \end{array} \right.
\end{equation}

Let us further assume that the function $F$
in Eq.~(\ref{defwF}) does not depend on the coordinates $\theta, \phi$.
We then obtain from (\ref{lambda})
\begin{equation}
\partial_z^3 \lambda = 0\,,
\end{equation}
which leads to
\begin{equation}
w = \frac{2i}{l}\frac{az^2 + bz + c}{2az + b}\,,
\end{equation}
where $a,b,c$ are complex integration constants. Now we shift
\begin{equation}
z \to z - \frac{a\bar b + b \bar a}{4a\bar a}\,,
\end{equation}
and define the real constants
\begin{eqnarray*}
n &=& i\left(\frac{\bar b}{4\bar a} - \frac{b}{4a}\right)\,, \\
Q &=& \frac{i}{2l}\left(\frac ca - \frac{\bar c}{\bar a}\right)\,, \\
P &=& -\frac{1}{2l}\left(2n^2 + \frac ca + \frac{\bar c}{\bar a}\right)\,.
\end{eqnarray*}
This yields
\begin{equation}
f = -\frac nl + \frac{Qz - nP}{z^2 + n^2}\,, \qquad
g = -\frac zl + \frac{Pz + nQ}{z^2 + n^2}\,,
\end{equation}
and
\begin{equation}
  -Nl^2 = \frac{z^4 + 2z^2(n^2 - lP) - 4nlQz + n^2(n^2 + 2lP)
  + l^2(Q^2 + P^2)}{z^2 + n^2}\,.
        \label{lapse}
\end{equation}
The shift vector $\omega_i$ can now be determined from
(\ref{omegaz}) and (\ref{omegaij}), with the result $\omega_{\theta} = 0$ and
\begin{equation}
\omega_{\phi} = \left\{\begin{array}{l@{\,,\quad}l}
                     -2c n \theta & k=0 \\ 2c n \cos\theta & k = 1 \\
                     -2c n \cosh\theta & k = -1\,, \label{omegaphi}
                     \end{array} \right.
\end{equation}
where $c$ denotes an integration constant that can be set equal to one by
rescaling $z \to \gamma z$, $t \to t/\gamma$, $n \to \gamma n$,
$Q \to \gamma^2 Q$, $P \to \gamma^2 P$, with $\gamma^{-2} = c$. The metric
finally reads 
\begin{equation}
  ds^2 = N(dt + \omega_{\phi} d\phi)^2 - \frac{dz^2}{N}
  + (z^2 + n^2) l^{-2} C_{ij} dx^i dx^j\,,
       \label{RNTNAdS}
\end{equation}
with $C_{ij}$, $N$ and $\omega_{\phi}$ given in Eqns.~(\ref{Cij}),
(\ref{lapse}) and (\ref{omegaphi}) respectively. The electromagnetic field
strength is easily obtained from (\ref{F}). One checks that the final solution
belongs to the Reissner-Nordstr\"om-Taub-Nut-AdS class of solutions
\cite{Alonso-Alberca:2000cs}, with arbitrary nut charge $n$ and electric
charge $Q$, whereas the magnetic charge $P$ and mass parameter $M$ are given by
\begin{equation}
  P = -\frac{kl^2 + 4n^2}{2l}\,, \qquad M = \frac{2nQ}{l}\,. \label{susycond}
\end{equation}

These are exactly the conditions on $P$ and $M$ found in
\cite{Alonso-Alberca:2000cs}, under which the RN-TN-AdS solutions preserve one
quarter of the supersymmetry\footnote{The other sign for $P$ given in
  \cite{Alonso-Alberca:2000cs} can be obtained by replacing $\phi \to -\phi$,
$n \to -n$.}.
Note that the integrability conditions (\ref{integrability}) are already
satisfied by our solution (\ref{RNTNAdS}), (\ref{susycond}), because they yield
exactly the quantization of the magnetic charge given in (\ref{susycond}).

It would be interesting to recover also the rotating BPS black holes studied
in \cite{Caldarelli:1998hg}.

\subsection{New BPS Solutions of Petrov Type $I$}
All previously known timelike supersymmetric solutions of $N=2$,
$D=4$ gauged supergravity were included in the Pleba\'nski-Demia\'nski class
of solutions, which are of Petrov type $D$ (or $O$). To show that new
supersymmetric geometries arise from the most general solution, we will now
explicitly extract a BPS solution of Petrov type I.

To simplify the equations, we restrict ourselves to $g=0$ solutions, and define
a new function $H\equiv1/f$. The equations in subsection \ref{constrbos} can
then be easily solved;
it turns out that $H$ can be chosen to be an arbitrary function of $x$ and $y$
such that $\Delta H\neq0$, and the solution reads
\begin{equation}
  \dd s^2=-\frac1{H^2}\left(\dd t+\om_i\,\dd x^i\right)^2
  +H^2\,\dd z^2+\frac{\ell^2\Delta H}{4H}\left(\dd x^2+\dd y^2\right)\,,
\end{equation}
where $\om_i$ is any solution of the curl equation
\begin{equation}
\p_x\om_y-\p_y\om_x=\ell\Delta H\,,
\end{equation}
and the corresponding electromagnetic field reads
\begin{equation}
  {\mathcal F}=-\p_i\left(\frac1H\right)\left(\dd t+\om_j\,\dd x^j\right)
  \wedge\dd x^i+\frac{\ell\Delta H}{4H}\,\dd x\wedge\dd y\,.
\end{equation}
The integrability condition (\ref{twist}) reduces then to the partial
differential equation
\begin{equation}
\Delta\ln\Delta H=6\frac{\Delta H}H-3\frac{\left(\nabla H\right)^2}{H^2}\,.
\end{equation}
It is easy to find a particular solution of these equations, given by
\begin{equation}
H=\frac{x^2}{\ell^2}
\end{equation}
and
\begin{equation}
\om=\frac{2x}\ell\,\dd y\,.
\end{equation}
This yields the BPS solution
\begin{equation}
  \dd s^2=-\frac{\ell^4}{x^4}\left(\dd t+\frac{2x}\ell\,\dd y\right)^2
  +\frac{x^4}{\ell^4}\,\dd z^2+\frac{\ell^2}{2x^2}
  \left(\dd x^2+\dd y^2\right)\,,
\end{equation}
\begin{equation}
  {\mathcal F}=\frac{2\ell^2}{x^3}\,\dd t\wedge\dd x
  -\frac{7\ell}{2x^2}\,\dd x\wedge\dd y\,.
\end{equation}
This metric has four Killing vectors,
\begin{equation}
T=\der{}{t}\,,\quad
Y=\der{}{y}\,,\quad
Z=\der{}{z}\,,\quad
K=t\der{}{t}-z\der{}{z}+\frac x2\der{}{x}+\frac y2\der{}{y}\,,
\end{equation}
acting transitively on the whole spacetime. Hence the solution represents a
homogeneous, stationary and geodesically complete BPS spacetime endowed with a
nonnull electromagnetic field.
A computation of the Weyl scalars of this metric shows that its Petrov type
is~$I$.
In the context of Einstein-Maxwell theory, this solution was obtained in
\cite{Ozsvath}.


\section{The Lightlike Case}\label{lightlike}

\subsection{Construction of the Bosonic Fields}

In the case when the Killing vector constructed from the Killing spinor
is lightlike, the algebraic constraints (\ref{fVfA}) -- (\ref{PhiPhi})
simplify to

\begin{equation}
V^2=A^2=f=g=\Phi^2=0\,,
\label{nullconstr1}
\end{equation}
\begin{equation}
V^{\mu}\Phi_{\mu\nu}=A^{\mu}\Phi_{\mu\nu}=0\,, \qquad
A\wedge\Phi=V\wedge\Phi=0\,, \qquad \Phi\wedge\Phi=0\,,
\label{nullconstr2}
\end{equation}
\begin{equation}
V\wedge A=0\,, \qquad\Phi_{(\mu}{}^\rho\eps_{\nu)\rho\al\be}=0\,.
\label{nullconstr3}
\end{equation}

The differential constraints read
\begin{equation}
{\mathcal F}_{\mu\nu}V^\nu=0\,,
\label{eqnullf}
\end{equation}
\begin{equation}
\frac12\eps_{\mu\nu\rho\si}V^\nu{\mathcal F}^{\rho\si}= -\frac1\ell A_\mu\,,
\label{eqnullg}
\end{equation}
\begin{equation}
  \nabla_\mu V_\nu=\frac1\ell\Phi_{\mu\nu}\,,
\label{eqnullV}
\end{equation}
\begin{equation}
  \nabla_\mu A_\nu=
  -{\mathcal F}_{(\mu}{}^\rho\eps_{\nu)\rho\al\be}\Phi^{\al\be} 
  +\frac14g_{\mu\nu}\eps_{\rho\si\al\be}{\mathcal F}^{\rho\si}\Phi^{\al\be}\,,
\label{eqnullA}
\end{equation}
\begin{equation}
  \nabla_\mu\Phi_{\al\be}=\frac2\ell g_{\mu[\al}V_{\be]}
  +2{\mathcal F}_{[\al}{}^\si\eps_{\be]\si\mu\nu}A^\nu
  +{\mathcal F}_\mu{}^\si\eps_{\si\al\be\nu}A^\nu
  +g_{\mu[\al}\eps_{\be]\nu\rho\si}A^\nu{\mathcal F}^{\rho\si}\,.
\label{eqnullPhi}
\end{equation}

From (\ref{nullconstr2}) and (\ref{eqnullV}) we get $V\wedge\dd V=0$,
hence $V$ is hypersurface orthogonal, and there exist two functions $H$ and
$u$ such that $V=H^{-1}\ \dd u$.
Furthermore one has
$V^{\mu}\nabla_{\mu} V_\nu = \frac{1}{\ell}V^{\mu}\Phi_{\mu\nu} = 0$,
so that $V$ is tangent to an affinely parametrized congruence of geodesics in
the surfaces of constant $u$. Let $v$ be the affine parameter,
$V=\frac\partial{\partial v}$.
Now use $u$ and $v$ as coordinates and choose coordinates $y^m$, $m=1,2$
for the transverse space. The most general metric reads
\begin{equation}
    \dd s^2=\frac1H\left({\mathcal G}\ \dd u^2+2\ \dd u\dd v\right)
    + H^{2\al}\gamma_{mn}\ \dd y^m\dd y^n\,,
\end{equation}
with $\al$ a real parameter to be chosen for convenience.
As $V$ is a Killing vector,$H$, $\mathcal G$ and $\ga_{mn}$ are functions of
$(u,y^m)$ only.
Let us introduce a null basis $(e^+,e^-,e^i)$ for the tangent space,
given by
\begin{equation}
    e^+=V=\frac1H\,\dd u\,, \qquad
    e^-=\dd v+\frac12{\mathcal G}\,\dd u\,, \qquad
    e^i= H^\al\,\check e^i\,,
\end{equation}
with $\check e^i=\check e^i{}_m\ \dd y^m$, $\delta_{ij}\check e^i\check e^j
=\gamma_{mn}\dd y^m\dd y^n$. From (\ref{eqnullV}) we obtain then
$\displaystyle \Phi=-\frac\ell{2H}\,\dd H\wedge e^+$.
The equation $V\wedge A=0$ yields $A=k(u, y^m)V=k(u, y^m)e^+$.
The function $k(u, y^m)$ can then be determined from the condition
$\dd A = 0$ and the definition of $e^+$. In this way one obtains
$k(u, y^m) = \kappa(u) H(u, y^m)$, where $\kappa(u)$ is a function of $u$
only. By a reparametrization $u=u(u')$ of the coordinate
$u$, it is possible to set $\kappa$ to a constant. The only two inequivalent
solutions are $\kappa=0$ (hence $A=0$) and $\kappa=1$:
\begin{equation}
    A=\kappa H(u,y^m)\,e^+\,,\qquad \kappa = 0, 1\,.
      \label{AkappaH}
\end{equation}
The transverse manifold is two-dimensional, hence conformally flat.
It is then possible to choose coordinates $x^i=x^i(u,y^m)$, $i=1,2$, such
that $\gamma_{mn}\,\dd y^m\dd y^n=\Omega^2(u,x^1,x^2)
\left[(\dd x^1)^2+(\dd x^2)^2\right]$.\\
Note that this coordinate transformation introduces mixed terms
$\dd u\dd x^i$.
In the coordinates $(u,v,x^i)$, the metric reads
\begin{equation}
    \dd s^2=\frac1H\left({\mathcal G}\ \dd u^2+2\ \dd u\dd v\right)
    + H^{2\al}\Omega^2\left(\dd x^i+a^i\,\dd u\right)
    \left(\dd x^i+a^i\,\dd u\right)\,,
    \label{metricuvxi}
\end{equation}
where $H$, $\mathcal G$, $\Omega$ and $a^i$ are functions of $(u,x^i)$.
The vierbein is now given by
\begin{equation}
    e^+=V=\frac1H\,\dd u,\quad
    e^-=\dd v+\frac12{\mathcal G}\,\dd u,\quad
    e^i= H^\al\Omega\left(\dd x^i+a^i\,\dd u\right)\,.
\end{equation}
Finally we choose an orientation such that $\eps_{+-12}=\eta$ with
$\eta^2=1$.

In order to determine the Maxwell field, we use
$V^{\mu}{\mathcal F}_{\mu\nu} = 0$, which yields
\begin{equation}
  {\mathcal F}={\mathcal F}_{+i}e^+\wedge e^i
  +\frac12{\mathcal F}_{ij}e^i\wedge e^j\,.
\end{equation}
The dual field is
\begin{equation}
\dual{\mathcal F}=
  \eta{\mathcal F}_{12}\,e^+\wedge e^-
  -\eta\ep_{ij}{\mathcal F}_{+j}\, e^+\wedge e^i\,,
\end{equation}
where we have defined $\ep_{12}=-\ep_{21}=1$. From (\ref{eqnullg}) and
(\ref{AkappaH}) one gets then ${\mathcal F}_{12}=-\frac{\eta\kappa}\ell H$,
and finally
\begin{equation}
    {\mathcal F}={\mathcal F}_{+i}e^+\wedge e^i
    -\frac{\eta\kappa}\ell H\,e^1\wedge e^2\,,
\end{equation}
\begin{equation}
    \dual{\mathcal F}=-\frac\kappa\ell H\,e^+\wedge e^-
    -\eta\ep_{ij}{\mathcal F}_{+j}\,e^+\wedge e^i\,.
\end{equation}
The Bianchi identity $\dd F=0$ and the Maxwell equation
$\dd\dual{\mathcal F}=0$ imply
\begin{equation}
    \partial_i\left(\Omega H^{\al-1}\ep_{ij}{\mathcal F}_{+j}\right)
    =-\frac{\eta\kappa}\ell\left[
      \partial_u\left(\Omega^2H^{2\al+1}\right)
      -\partial_i\left(\Omega^2H^{2\al+1}a^j\right)
      \right]
      \label{bianchiid}
\end{equation}
and
\begin{equation}
    \partial_i\left(\Omega H^{\al-1}{\mathcal F}_{+i}\right)=0
    \label{sourceeq}
\end{equation}
respectively.
The general solution of (\ref{sourceeq}) is
\begin{equation}
    {\mathcal F}_{+i}=\Omega^{-1}H^{-\al+1}\ep_{ij}\partial_j\psi\,,
    \label{F+i}
\end{equation}
where the function $\psi(u, x^i)$ is determined by the Bianchi identity
(\ref{bianchiid}), which yields
\begin{equation}
    \Delta\psi
    =\frac{\eta\kappa}\ell\left[
      \partial_u\left(\Omega^2H^{2\al+1}\right)
      -\partial_i\left(\Omega^2H^{2\al+1}a^i\right)
      \right]\,, \label{Deltapsi}
\end{equation}
where $\Delta\equiv\partial_1^2+\partial_2^2$ is the flat transverse space
Laplacian.

We come now to the differential equation (\ref{eqnullA}) for $A$.
As we have
\begin{equation}
\eps_{\rho\sigma\alpha\beta}{\mathcal F}^{\rho\sigma}\Phi^{\alpha\beta}=0\,,
\end{equation}
it simplifies to
\begin{equation}
\nabla_\mu A_\nu=-{\mathcal F}_{(\mu}{}^\rho\eps_{\nu)\rho\al\be}\Phi^{\al\be}\,.
                 \label{nablaA}
\end{equation}
Using (\ref{AkappaH}) and (\ref{eqnullV}) yields
\begin{equation}
\nabla_\mu A_\nu=\kappa V_\nu\partial_\mu H+\frac\kappa\ell H\Phi_{\mu\nu}\,.
\end{equation}
Comparing this with (\ref{nablaA}), one obtains
in the orthonormal frame
\begin{equation}
    \kappa V_a\partial_\mu H=\frac\kappa\ell H e^b{}_\mu{\mathcal F}_{(a}{}^c
    \left(\dual\Phi\right)_{b)c}\,.
\end{equation}
The only non trivial statement comes from the components $a=+$ and $\mu=u$,
\begin{equation}
    \kappa\partial_uH=\left(\kappa a^k+\frac{\eta\ell}{\Omega^2H^{2\al+1}}
    \partial_k\psi\right)\partial_k H\,.
    \label{nontriv}
\end{equation}

In order to solve the differential equation (\ref{eqnullPhi}) for $\Phi$,
we use (\ref{eqnullV}) to get
\begin{equation}
\nabla_\mu\Phi_{\al\be}=\ell\nabla_\mu\nabla_\al V_\be\,.
\end{equation}
Since $V$ is a Killing vector, we have
\begin{equation}
\nabla_\mu\nabla_\nu V_\rho=-R_{\nu\rho\mu}{}^{\sigma}V_\sigma\,.
\end{equation}
In the orthonormal frame, (\ref{eqnullPhi}) becomes then
\begin {equation}
    -\ell R_{abc}{}^+=\frac2\ell\eta_{c[a}V_{b]}
    +2{\mathcal F}_{[a}{}^d\eps_{b]dce}
    A^e+{\mathcal F}_c{}^d\eps_{dabe}A^e
    +\eta_{c[a}\eps_{b]def}A^d{\mathcal F}^{ef}\,.
\end {equation}
The independent components of this equation are
\begin{equation}
    R_{+-+-}=\frac1{\ell^2}\left(1-\kappa H^2\right)\,,
\end{equation}
\begin{equation}
    R_{+-+i}=\frac{2\eta\kappa}{\ell\Omega H^{\al-1}}\partial_i\psi\,,
\end{equation}
\begin{equation}
    R_{+-ij}=\frac{2\kappa}{\ell^2}H^2\ep_{ij}\,, \label{R+-ij}
\end{equation}
\begin{equation}
    R_{+i-j}=-\frac1{\ell^2}\left(1+\kappa H^2\right)\delta_{ij}\,,
\end{equation}
\begin{equation}
    R_{+--i}=R_{-i-j}=R_{-ijk}=0\,. \label{R+--i}
\end{equation}
The equations (\ref{R+--i}) are automatically satisfied
by the Riemann tensor (\ref{Riem+-}), (\ref{Riem-i}) of the metric
(\ref{metricuvxi}).

From (\ref{R+-ij}) we have $\frac{2\kappa}{\ell^2}H^2\ep_{ij}=0$,
and thus $\kappa = 0$, because $H = 0$ would lead to a degenerate
metric.
The remaining equations then simplify to
\begin{equation}
    \frac{\partial_i H\partial_i H}{4\Omega^2H^{2\al+2}}=1/\ell^2\,,
    \label{eq1}
\end{equation}
\begin{equation}
      a^k\partial_k\partial_i\left(\ln H\right)
      -\partial_u\partial_i\left(\ln H\right)
      +\partial_i\left(\ln H\right)\partial_u\left(\ln\Omega H^\al\right)
      +\partial_{[i}a_{j]}\partial_j\left(\ln H\right)
      =0\,, \label{eq2}
\end{equation}
\begin{eqnarray}
    & &
    \frac1{2\Omega^2 H^{2\al}}
    \left[\frac1H\partial_i\partial_jH
      -2\partial_{(i}\ln H \partial_{j)}\ln\Omega
      -\left(2\al+\frac32\right)\partial_i\ln H \partial_j\ln H
      \right.\nonumber\\
      & &
      \left. \qquad\qquad\qquad\phantom{\frac32}
      +\delta_{ij}\partial_k\left(\ln H\right)\partial_k\left(\ln\Omega
    H^\al\right) \right]=-\frac1{\ell^2}\delta_{ij}\,,
    \label{eq3}
\end{eqnarray}
in addition to $\Delta\psi=0$ following from (\ref{Deltapsi}) and
$\partial_k\psi\partial_k H=0$ from (\ref{nontriv}).

The Einstein equations are almost automatically satisfied if the
fields solve the previous conditions for the existence of a Killing spinor.
The only component that has to be verified explicitly is the $(+,+)$
one (cf.~appendix \ref{integrab} and ref.~\cite{Gauntlett:2002nw}).
The Einstein equations read
\begin{equation}
    G_{\mu\nu}-\frac3{\ell^2}g_{\mu\nu}=
    2\left({\mathcal F}_\mu{}^\rho {\mathcal F}_{\nu\rho}
    -\frac14{\mathcal F}^2g_{\mu\nu}\right)\,,
\end{equation}
whose $(+,+)$ component yields
\begin{equation}
    R_{++}=2{\mathcal F}_{+i}{\mathcal F}_{+i}
    =\frac{2\partial_i\psi\partial_i\psi}{\Omega^2H^{2\al-2}}\,,
\label{eq++}\end{equation}
where we have substituted ${\mathcal F}_{+i}$ from (\ref{F+i}).
The component $R_{++}$ of the Ricci tensor is given in (\ref{ricci++}).

We have thus to solve the equations (\ref{eq1})--(\ref{eq3}), the $(+,+)$
component of the Einstein equation, and the equations for $\psi$.

The trace of Eqn.~(\ref{eq3}) yields
\begin{equation}
  \frac1H\Delta H-\frac32\frac{\left(\nabla H\right)^2}{H^2}
  =-\frac4{\ell^2}\Omega^2H^{2\al}\,.
\end{equation}
Using Eqn.~(\ref{eq1}) to eliminate $\Omega$, one obtains
\begin{equation}
    \Delta H-\frac1{2H}\left(\nabla H\right)^2=0\,.
\end{equation}
This equation is equivalent to $\Delta H^{1/2}=0$, hence $H^{1/2}$ solves the
flat Laplace equation on the $(x,y)$ plane.
We introduce now complex coordinates $\zeta=x^1+ix^2$, $\bar\zeta=x^1-ix^2$ and
the holomorphic and anti-holomorphic derivative operators
$\p=\frac12\left(\p_1-i\p_2\right)$,
$\bar\p=\frac12\left(\p_1+i\p_2\right)$.
The equation for $H$ now reads $\p\bar\p H^{1/2}=0$, and the general solution
is the sum of a holomorphic and an anti-holomorphic function,
\begin{equation}
H^{1/2}=h(u, \ze)+\tilde{h}(u,\bar\ze)\,.
\end{equation}
From the reality of $H$ we have
\begin{equation}
\tilde{h}(u,\ze)=\overline{h(u,\ze)}=\bar h(u,\bar\ze)\,.
\end{equation}
Finally,
\begin{equation}
    H(u,\ze,\bar\ze)=\left[h(u,\ze)+\bar h(u,\ze)\right]^2\,,
    \label{Hhbarh}
\end{equation}
where $h(u,\ze)$ is an arbitrary function.

The conformal factor $\Omega$ can now be determined using again
Eqn.~(\ref{eq1}), with the result
\begin{equation}
    \Omega(u,\ze,\bar\ze)=\frac{2\ell}{\left(h+\bar h\right)^\al}
    \left(\p h\bar\p\bar h\right)^{1/2}\,. \label{resOmega}
\end{equation}
The component $(1,2)$ of Eqn.~(\ref{eq1}) reduces to
\begin{equation}
    (\al+1)\left[\left(\p h\right)^2-\left(\bar\p\bar h\right)^2\right]=0\,.
    \label{comp12}
\end{equation}
We can, with no loss of generality, set $\al=-1$, so that (\ref{comp12})
imposes no further restriction on the arbitrary function $h$.

Using the expressions (\ref{Hhbarh}) for $H$ and
(\ref{resOmega}) for $\Omega$, the metric reads
\begin{equation}
  \dd s^2=\frac1{(h+\bar h)^2}\left({\mathcal G}\,\dd u^2+2\,\dd u\dd v
  + 4\ell^2\p h\bar\p\bar h\,\left|\dd\ze+a\,\dd u\right|^2\right)\,.
\end{equation}
We can define new coordinates $\xi=2\ell h(\ze,u)$ and
$\bar\xi=2\ell\bar h(\ze,u)$, to obtain
\begin{equation}
  \dd s^2=\frac{4\ell^2}{(\xi+\bar \xi)^2}\left({\mathcal G}\,\dd u^2
  +2\,\dd u\dd v + \left|\dd\xi+b\,\dd u\right|^2\right)\,,
\end{equation}
where the function $b$ is defined by $b=-2\ell\left(\p_u-a^k\p_k\right)h$.
This is exactly a metric of the original form, with $h(\ze,u)=\ze/2\ell$ and
$a=b$. Thus, without loss of generality, we can restrict to solutions with
\begin{equation}
H=\left(\frac{\ze+\bar\ze}{2\ell}\right)^2\,,\qquad
\Omega=\frac{\ze+\bar\ze}{2\ell}\,.
\end{equation}
In the coordinates $x^k=(x,y)$ (recall $\ze=x+iy$), the metric reads
\begin{equation}
  \dd s^2=\frac{\ell^2}{x^2}\left({\mathcal G}\,\dd u^2+2\,\dd u\dd v
  +\left(\dd x^k+a^k\,\dd u\right)\left(\dd x^k+a^k\,\dd u\right)\right)\,.
\end{equation}
In order to determine $a^k$ we have to solve Eqn.~(\ref{eq2}), which simplifies
to
\begin{equation}
a^x=0\,,\qquad
\p_xa^y=0\,.
\end{equation}
Hence $a^y=a(u,y)$ is an arbitrary function of $u$ and $y$.\\
The electromagnetic field is determined by the scalar field $\psi$, restricted
by $\Delta\psi=0$, $\nabla\psi\cdot\nabla H=0$. The second equation implies
$\p_x\psi=0$, hence $\psi=\psi(u,y)$. Then the first one imposes
$\p_y^2\psi=0$, which is satisfied by
\begin{equation}
\psi=y\varphi'(u)+\psi_0(u)\,,
\end{equation}
where $\varphi$ and $\psi_0$ are two arbitrary functions of $u$.
The resulting electromagnetic field is
${\mathcal F}=\varphi'(u)\,\dd u\wedge\dd x$, and does not depend on $\psi_0$,
hence we can take $\psi_0=0$ in the following.

Now the metric reads
\begin{equation}
  \dd s^2=\frac{\ell^2}{x^2}\left({\mathcal G}(u,x,y)\,\dd u^2+2\,\dd u\dd v
  +\dd x^2+\left(\dd y+a(u,y)\,\dd u\right)^2\right)\,,
\label{sola}\end{equation}
where the function $\mathcal G$ must satisfy the $(++)$ component of Einstein's
equations,
\begin{equation}
  \Delta{\mathcal G}-\frac2x\partial_x{\mathcal G}=-\frac{4x^2}{\ell^2}
  \left(\varphi'\right)^2
  +2\p_y\p_ua-2\p_y\left(a\p_ya\right)\,.
\end{equation}
This equation reduces to the Siklos equation \cite{Siklos:1985} when
$\varphi=a=0$.

We can eliminate $a(u, y)$ by shifting $v \mapsto v+g(u,x,y)$, with $g$ an
arbitrary function of its arguments. The metric remains of the same form, and
the functions $\mathcal G$ and the vector $a^k$ transform according to
\begin{equation}
  {\mathcal G}\mapsto{\mathcal G}'={\mathcal G}+2g_{,u}-2a\cdot\nabla g
  -\left(\nabla g\right)^2\,,
\end{equation}
\begin{equation}
  a^k\mapsto a'^k=a^k+\p^kg\,.
\end{equation}
Thus, taking $g(u,x,y)=-\int\!a(u,y)\,\dd y$ in (\ref{sola}), we can
eliminate the vector $a^k$ and, up to diffeomorphisms, the most
general solution has $a=0$.

\subsection{Integrability Conditions and Unbroken Supersymmetry Generators}

We finally come to the unbroken supersymmetry generators.
First of all, we have $f=g=0$, hence $V^\mu\Ga_\mu\ep=i(f+ig\Ga_5)\ep=0$, and
given $V=e^+$ it follows that
\begin{equation}
\Ga^+\ep=0\,,
\end{equation}
so half of the supersymmetries is broken.
The $v$ component of the Killing spinor equation yields
\begin{equation}
\frac{\p\ep}{\p v}=0\,,
\end{equation}
so that $\epsilon$ is independent of $v$.
The $y$ component reads
\begin{equation}
\left[\p_y+\frac1{2x}\Ga_2\left(1-\Ga_1\right)\right]\ep=0\,,
\end{equation}
and is solved by taking
\begin{equation}
\frac{\p\ep}{\p y}=0\,,\qquad
\left(1-\Ga_1\right)\ep=0\,.
\end{equation}
For such a spinor, the remaining equations ($u$ and $x$ components) reduce to
\begin{equation}
\left(\p_x-\frac i\ell\varphi+\frac1{2x}\Ga_1\right)\ep=0\,,\qquad
\left(\p_u-\frac{ix}\ell\varphi'\Ga_1\right)\ep=0\,.
\end{equation}
This system admits the solution
\begin{equation}
\ep=\sqrt{\frac\ell x}\exp{\left(\frac{ix}\ell\varphi(u)\right)}\ep_0\,,
\end{equation}
where $\ep_0$ is an arbitrary constant spinor satisfying
\begin{equation}
\Ga_-\ep_0=0\,,\qquad
\left(1-\Ga_1\right)\ep_0=0\,.
\end{equation}
Hence the lightlike supersymmetric solutions are one quarter BPS states.

\subsection{The General Lightlike BPS Solution}
In conclusion, the general lightlike BPS solution is an electrovac travelling
wave given by an arbitrary function $\varphi(u)$ and a solution
${\mathcal G}(u,x,y)$ of the equation
\begin{equation}
  \Delta{\mathcal G}-\frac2x\partial_x{\mathcal G}=-\frac{4x^2}{\ell^2}
  \left(\varphi'\right)^2\,.
\end{equation}
The metric reads
\begin{equation}
  \dd s^2=\frac{\ell^2}{x^2}\left[{\mathcal G}(u,x,y)\,\dd u^2+2\,\dd u\dd v
  +\dd x^2+\dd y^2\right]\,,
\end{equation}
and the
null
electromagnetic field is given by
\begin{equation}
  {\mathcal F}=\dd{\mathcal A} =\varphi'(u)\,\dd u\wedge\dd x\,,\qquad
  {\mathcal A}=\varphi(u)\,\dd x\,.
\end{equation}
This family of solutions has a Virasoro symmetry with non-zero central charge
as in the pure gravitational case \cite{Banados:1999tw}, corresponding to the
reparameterization freedom in the coordinate $u$, $u=f(u')$.
Defining a new complex coordinate $\xi$ by
\begin{equation}
x+iy=\ell\,\frac{\xi-\ell}{\xi+\ell}\,,
\end{equation}
and performing the change of coordinates, we see that these solutions form the
$(IV)_0$ family of metrics found in \cite{Ozsvath:qn}, describing exact
gravitational and electromagnetic waves of arbitrary profile propagating in
AdS space along the direction $v$ \cite{Podolsky:2002sy}.
The special case
\begin{equation}
  {\mathcal G}=\frac{P^2}{\ell^4}x^3-\frac{P_e^2}{\ell^6}x^3
\end{equation}
is a charged Kaigorodov space, obtained in \cite{Cai:2003xv} as the
ultrarelativistic limit of a boosted non-extremal charged domain wall.

\section{Maximal Supersymmetry}

\label{maxsusy}

In order to get a maximally supersymmetric solution, we require that
the integrability condition (\ref{intcond}) following from the Killing
spinor equation imposes no algebraic constraints on the Killing spinor.
This means that the terms which are zeroth, first, second, third and fourth order
in the gamma-matrices must vanish independently. From the zeroth order term we immediately
obtain ${\mathcal F} = 0$\footnote{Note that this zeroth order term comes with prefactor
$1/\ell$, so that it is absent in the ungauged case, where maximally supersymmetric
solutions with nonvanishing gauge field exist.}. Using this, the integrability condition
simplifies to
\begin{equation}
R_{\alpha\beta\nu\mu} = \frac{1}{\ell^2}(g_{\alpha\nu} g_{\beta\mu} - g_{\alpha\mu}
                                      g_{\beta\nu})\,,
\end{equation}
so that the spacetime has constant curvature. We conclude that the only maximally
supersymmetric geometry is given by AdS$_4$ with vanishing gauge field. This is
analogous to the five-dimensional case \cite{Gauntlett:2003fk}.



\section{Final remarks}

\label{final}

We conclude this paper by pointing out some possible extensions of the work
presented here. First of all, it would be interesting to uplift e.~g.~the new
lightlike solutions to eleven dimensions and to study their M-theory
interpretation.

Furthermore, one could refine our classification in the sense of finding
the additional restrictions on the geometries in order that they preserve
more than one supersymmetry, and to see whether e.~g.~$3/4$ supersymmetric
solutions are possible.

Although electromagnetic duality invariance is broken in the gauged
theory due to the minimal coupling of the gravitini to the graviphoton,
a generalized duality invariance was discovered in the supersymmetric
subclass of the Pleba\'nski-Demia\'nski solution, which rotates also the mass
parameter into the nut charge and vice-versa \cite{Alonso-Alberca:2000cs}.
It would be interesting to see whether this duality can be understood within
our formalism.

Finally, one could consider matter-coupled gauged supergravity, where a large
class of supersymmetric black holes is known
\cite{Sabra:1999ux,Chamseddine:2000bk}, and see whether
a classification of BPS solutions is still feasible.
Work in these directions is in progress.

\section*{Acknowledgements}
\small

This work was partially supported by INFN, MURST and
by the European Commission RTN program
HPRN-CT-2000-00131, in which M.~M.~C.~and D.~K.~are
associated to the University of Torino.
\normalsize

\newpage

\begin{appendix}

\section{Conventions}
Throughout this paper, the conventions are as follows:
$a,b,\ldots$ refer to $D=4$ tangent space indices, and $\mu,\nu,\ldots$
refer to $D=4$ world indices. The signature is $(-,+,+,+)$,
$\eps_{0123}=+1$.

The gamma matrices are defined to satisfy the four-dimensional Clifford algebra
$\{\Ga_a,\Ga_b\}=2\eta_{ab}$, and the parity matrix is $\Ga_5 = i\Ga_{0123}$.
We antisymmetrize with unit weight, i.~e.~$\Ga_{ab} \equiv
\Ga_{\left[a\right.}\Ga_{\left.b\right]} \equiv \frac{1}{2}[\Ga_a,\Ga_b]$ etc.
The Dirac conjugate is defined by $\bar\psi=i\psi^\dagger\Ga^0$.

Finally, for two-dimensional submanifolds, we use $i,j,\ldots$ as $D=2$
indices. The indices can take the value $1$, $2$ and $\ep_{12}=+1$.
The geometric objects associated to the bidimensional metric $h_{ij}$ are
surmounted by a check sign; hence $\check\omega^{12}{}_i$ is the spin
connection, and the curvature tensor reads
$\check{\mathcal R}^{12}=\check\dd\check\om^{12}$, where 
$\check\dd=\dd x\,\p_x+\dd y\,\p_y$ is the two-dimensional exterior
derivative. We denote by
\begin{equation}
\Delta_h\equiv\frac1{\sqrt{h}}\p_i\left(\sqrt{h}h^{ij}\p_j\right)
\end{equation}
the Laplacian associated to $h_{ij}$, and with $\Delta$ the usual flat
bidimensional Laplacian.

\section{Decomposition of a two-form with respect to a vector $K$}
\label{decomp}
Given a two-form $F$ and a vector $K$, we can decompose $F$ in an
electric and a magnetic part, defined by
\begin{equation}
E=-i_KF=\dual\left(K\wedge\dual F\right)\,, \qquad
B=i_K\dual F=\dual\left(K\wedge F\right)\,.
\end{equation}
Then the two-form reads
\begin{equation}
F=-\frac1{K^2}\left(K\wedge E+\dual\left(K\wedge B\right)\right)\,.
\end{equation}
If $K$ is a Killing vector, it follows from the definition that
$\lie_KE=\lie_KB=0$.

\section{Useful Relations for a Spinor $\ep$}
\begin{equation}
  \bar\ep\Ga_\mu\Ga_\nu\ep=fg_{\mu\nu}-i\Phi_{\mu\nu}\,, \qquad
  \bar\ep\Ga_5\Ga_\mu\Ga_\nu\ep=-igg_{\mu\nu}+\frac12\eps_{\mu\nu\rho\si}
  \Phi^{\rho\si}\,,
\end{equation}
\begin{equation}
  \bar\ep\Ga_{\mu\nu}\Ga_\rho\ep=-\eps_{\mu\nu\rho\si}A^\si
  -2iV_{[\mu}g_{\nu]\rho}\,,\qquad
  \bar\ep\Ga_\mu\Ga_{\nu\rho}\ep=-\eps_{\mu\nu\rho\si}A^\si
  -2ig_{\mu[\nu}V_{\rho]}\,,
\end{equation}
\begin{equation}
  \bar\ep\Ga_5\Ga_{\mu\nu}\Ga_\rho\ep=-\eps_{\mu\nu\rho\si}V^\si
  -2iA_{[\mu}g_{\nu]\rho}\,,
\end{equation}
\begin{equation}
  \bar\ep\Ga_5\Ga_\mu\Ga_{\nu\rho}\ep=-\eps_{\mu\nu\rho\si}V^\si
  -2ig_{\mu[\nu}A_{\rho]}\,,
\end{equation}
\begin{equation}
  \bar\ep\Ga_{\mu\nu}\Ga_{\rho\si}\ep=-g\eps_{\mu\nu\rho\si}
  +2i\Phi_{\mu[\rho}g_{\si]\nu}
  -2ig_{\mu[\rho}\Phi_{\si]\nu}
  -2fg_{\mu[\rho}g_{\si]\nu}\,,
\end{equation}
\begin{equation}
  \bar\ep\Ga_5\Ga_{\mu\nu}\Ga_{\rho\si}\ep=-if\eps_{\mu\nu\rho\si}
  +\eps_{\rho[\mu}{}^{\al\be}g_{\nu]\si}\Phi_{\al\be}
  -\eps_{\si[\mu}{}^{\al\be}g_{\nu]\rho}\Phi_{\al\be}
  +2igg_{\mu[\rho}g_{\si]\nu}\,,
\end{equation}
\begin{equation}
  \bar\ep\Ga_\mu\Ga_{\rho\si}\Ga_\nu\ep=-g\eps_{\mu\nu\rho\si}
  -2i\Phi_{\mu[\rho}g_{\si]\nu}
  -2ig_{\mu[\rho}\Phi_{\si]\nu}
  -ig_{\mu\nu}\Phi_{\rho\si}
  +2fg_{\mu[\rho}g_{\si]\nu}\,,
\end{equation}
\begin{eqnarray}
&&  \bar\ep\Ga_5\Ga_\mu\Ga_{\rho\si}\Ga_\nu\ep=-if\eps_{\mu\nu\rho\si}
  -2igg_{\mu[\rho}g_{\si]\nu}
  +\eps_{\mu[\rho}{}^{\al\be}g_{\si]\nu}\Phi_{\al\be}\nonumber\\
&& \qquad\qquad\qquad\qquad
  +g_{\mu[\rho}\eps_{\si]\nu}{}^{\al\be}\Phi_{\al\be}
  +\frac12g_{\mu\nu}\eps_{\rho\si\al\be}\Phi^{\al\be}\,.
\end{eqnarray}

\section{Geometry of the Lightlike Solution}
Defining $\tilde\nabla\equiv\partial_u-a^k\partial_k$, 
the spin connection and the components of the Riemann and Ricci tensors needed
to solve the lightlike case read as follows.
\paragraph{Spin connection:}
\begin{equation}
  \omega_{+-}=-\frac{\partial_iH}{2\Omega H^{\al+1}}\,e^i\,, \qquad
  \omega_{-i}=-\frac{\partial_iH}{2\Omega H^{\al+1}}\,e^+\,,
\end{equation}
\begin{eqnarray}
  \omega_{+i}&=&\frac{\partial_i{\mathcal G}}{2\Omega H^{\al-1}}\,e^+
  -\frac{\partial_iH}{2\Omega H^{\al+1}}\,e^-\nonumber\\
  &&+\left(H\partial_{(i}a_{j)}-\frac1{\Omega H^{\al-1}}
  \tilde\nabla\left(\Omega H^\al\right)
  \delta_{ij}\right)\,e^j\,,
\end{eqnarray}
\begin{equation}
  \omega_{ij}=-H\partial_{[i}a_{j]}e^+
  +\frac2{\Omega^2H^{2\al}} \delta_{k[i}\partial_{j]}
  \left(\Omega H^\al\right)\,e^k\,.
\end{equation}

\paragraph{Riemann tensor:}
\begin{eqnarray}
  R_{+-}=\frac{\partial_i H\partial_i H}{4\Omega^2H^{2\al+2}}\,e^+\wedge e^-
  -\frac1{2\Omega H^{\al-1}}
  \left[
    \tilde\nabla\partial_i\left(\ln H\right)
    \right.\qquad\qquad\quad &&\nonumber\\
  \left.
    -\partial_i\left(\ln H\right)
    \tilde\nabla\left(\ln \Omega H^\al\right)
    +\partial_{[k}a_{i]}\partial_k\left(\ln H\right)
    \right]\,e^+\wedge e^i\,, \label{Riem+-}&&
\end{eqnarray}
\begin{eqnarray}
    R_{-i}=
    \frac1{2\Omega^2 H^{2\al}}
    \left[\frac1H\partial_i\partial_jH
    -2\partial_{(i}\left(\ln H\right)\partial_{j)}\left(\ln\Omega\right)
  \phantom{\frac32}\right.\qquad\qquad\qquad&&\nonumber\\
  \left.
    -\left(2\al+\frac32\right)\partial_i\left(\ln H\right)
    \partial_j\left(\ln H\right)
  \right.\qquad\qquad\qquad &&\nonumber\\
  \left.\phantom{\frac32}
    +\delta_{ij}\partial_{k}\left(\ln H\right)
    \partial_k\left(\ln\Omega H^\al\right)
    \right]\,e^+\wedge e^j\,.&& \label{Riem-i}
\end{eqnarray}

\paragraph{$(++)$ component of the Ricci tensor:}
  \begin{eqnarray}
    R_{++}=\frac1{2\Omega^2H^{2\alpha-1}}\left(-\Delta{\mathcal G}
    +\partial_k{\mathcal G}\partial_k\ln H\right)
    +H^2\left(-2\tilde\nabla\tilde\nabla\ln\Omega H^\al
    \right.\nonumber\\
    -2\tilde\nabla\ln\Omega H^{\al+1}\tilde\nabla\ln\Omega H^\al
    +\left(\partial_ka^k\right)\tilde\nabla\ln\Omega^2 H^{2\al+1}\nonumber\\
    \left.
    +\tilde\nabla\partial_ka^k-\frac12(\p_ia^j)(\p^ia_j)
    -\frac12(\p_ia^j)(\p_ja^i)\right)\,.
    \label{ricci++}
  \end{eqnarray}


\section{Integrability Conditions and Einstein Equations}

\label{integrab}

The Killing spinor equation (\ref{killspinequ}) implies the integrability condition
\begin{eqnarray}
[\sder_{\nu}, \sder_{\mu}]\epsilon &=& \left[\frac 1\ell(\dual{\mathcal F}_{\nu\mu}\Gamma_5
- i{\mathcal F}_{\nu\mu}) + \frac 1{2\ell^2}\Gamma_{\nu\mu} +
\frac 14{{\mathcal R}^{ab}}_{\nu\mu}\Gamma_{ab}\right. \nonumber \\
&& - {\mathcal F}^{\alpha\beta}{\mathcal F}_{\beta [\nu}\Gamma_{\mu]\alpha}
+ \frac 14 {\mathcal F}_{\alpha\beta}{\mathcal F}^{\alpha\beta}\Gamma_{\nu\mu}
- \frac i{\ell}{{\mathcal F}^{\alpha}}_{[\nu}\Gamma_{\mu]\alpha} \nonumber \\
&& \left.-\frac i2 \Gamma_{\alpha\beta[\nu}\nabla_{\mu]}{\mathcal F}^{\alpha\beta}
- i\nabla_{[\nu}{{\mathcal F}_{\mu]}}^{\alpha}\Gamma_{\alpha}\right]\epsilon = 0\,.
\label{intcond}
\end{eqnarray}
Contracting this with $\Gamma^{\mu}$ and using the Bianchi identity
$R_{\nu[\mu\alpha\beta]} = 0$, we obtain
\begin{equation}
E_{\nu\alpha}\Gamma^{\alpha}\epsilon + i\nabla_{\mu}{\mathcal F}^{\mu\alpha}
\Gamma_{\alpha}\Gamma_{\nu}\epsilon + \frac 12 \left[\nabla^{[\sigma}{\mathcal F}^{\alpha\beta]}
\epsilon_{\sigma\alpha\beta\nu}\Gamma_5 + 3i\nabla_{[\nu}{\mathcal F}_{\alpha\beta]}
\Gamma^{\alpha\beta}\right]\epsilon = 0\,,
\end{equation}
where we defined
\begin{equation}
E_{\nu\alpha} = R_{\nu\alpha} + \frac 3{\ell^2}g_{\nu\alpha} + 2 {\mathcal F}_{\nu\beta}
{{\mathcal F}^{\beta}}_{\alpha} + \frac 12 {\mathcal F}^2 g_{\nu\alpha}\,.
\end{equation}
Let us assume that the Maxwell equations and the Bianchi identity for ${\mathcal F}$ hold.
We then conclude that
\begin{equation}
E_{\nu\alpha}\Gamma^{\alpha}\epsilon = 0\,.
\end{equation}
If we multiply this from the left with $\bar\epsilon$ we get
\begin{equation}
E_{\nu\alpha}V^{\alpha} = 0\,. \label{firstequE}
\end{equation}
On the other hand, multiplying with $E_{\nu\sigma}\Gamma^{\sigma}$ yields
\begin{equation}
E_{\nu\sigma}{E_{\nu}}^{\sigma} = 0\,, \label{secondequE}
\end{equation}
where we do not sum over the index $\nu$. One can now proceed analogously to
\cite{Gauntlett:2002nw} and use (\ref{firstequE}) and (\ref{secondequE})
to show that in the timelike case the Killing spinor equations (plus the Maxwell equations
and Bianchi identity for ${\mathcal F}$) imply the
Einstein equations $E_{\nu\alpha} = 0$, whereas in the lightlike case the
component $E_{++} = 0$ must be additionally imposed.


\end{appendix}


\end{document}